# A C-band microwave rectifier without capacitors for microwave power transmission

BIAO ZHANG, WAN JIANG, CHENGYANG YU AND CHANGJUN LIU

*A microwave rectifier at 5.8 GHz without any capacitors is presented, which owns a measured MW-to-DC conversion efficiency of 68.1%. A harmonic rejection filter and a DC pass filter, which replace lumped capacitors in conventional microwave rectifiers, are applied to suppressing the harmonics produced by an HSMS-286 Schottky diode during rectifying. At the fundamental frequency, a microstrip impedance transformer which contains a shunt $\lambda_g/8$ short-ended microstrip transmission line and two short series microstrip transmission lines are applied to compensating the imaginary impedance of the diode and matching the input impedance of the rectifier. The measured MW-to-DC conversion efficiency agrees well to the simulated results. The novel rectifier without any lumped passive elements may be applied for power transmission system at higher microwave frequencies.*

## I. INTRODUCTION

A key component of a microwave power transmission (MPT) systems is the microwave rectifier, which converts microwave power into DC power and was firstly proposed by W.C. Brown in 1964 [1]. Owing to its excellent properties, the rectifying circuits have been widely applied in all kinds of recentennas [2–4], Radio frequency identification (RFID) tags [5], wireless sensors [6], and space energy-harvesting devices [7] after half century of development, which leads to great requirements on wireless power transmission. The MPT system, including microwave rectifiers, have been extensively studied at 5.8 GHz [8, 9], since the frequency is just within the ISM band. A microwave rectifier is composed of a DC-blocking circuit, an input filter, a rectifying diode, an output DC filter, and a DC load. Usually the DC-blocking circuit and output DC filter are realized with lumped elements, i.e. chip capacitors, respectively.

A novel microwave rectifier, as shown in Fig. 1, is presented in this paper. It consists of an input microstrip harmonic rejection filter, a Schottky diode, a $\lambda_g/8$ microstrip line, and an output microstrip DC pass filter. Because of the removement of the lumped capacitors, the microwave-rectifying circuit may find more applications in higher frequency microwave wireless power transmission systems.

School of Electronics and Information Engineering, Sichuan University, Chengdu, China. Phone: +86 28 8546 3882
**Corresponding author:**
C. Liu
Email: cjliu@ieee.org

## II. RECTIFYING CIRCUIT DESIGN

### A) Harmonic rejection filter

A compact microstrip harmonic rejection filter is introduced for the capacitor-less microwave rectifier. The low-pass filter with the cut-off frequency of 7.3 GHz is composed of one thin microstrip line and two microstrip patches with a dimensions of 8.1 mm × 3.6 mm on a F4B-2 substrate with $\varepsilon_r =$ 2.65 and thickness of 1 mm, as shown in Fig. 2. The thin line and the patches operate as a series inductor and two parallel capacitors, respectively. Figure 3 shows the measured results of the low-pass filter. The insertion losses of the filter are 0.3 dB at 5.8 GHz, 15.8 dB at 11.6 GHz, and 8.4 dB at 17.4 GHz, respectively. The fundamental frequency is located in the pass-band, and its harmonics are rejected to the rectifying diode by the harmonic rejection filter. The input and output impedances of the filter are 50 Ω to match the microwave source and load, respectively.

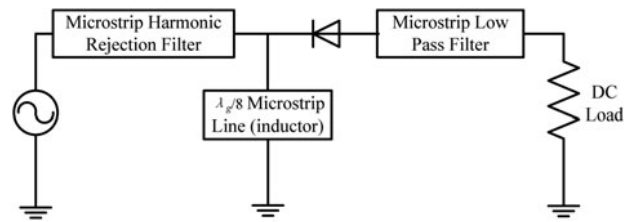

**Fig. 1.** Scheme of the proposed rectifying circuit.

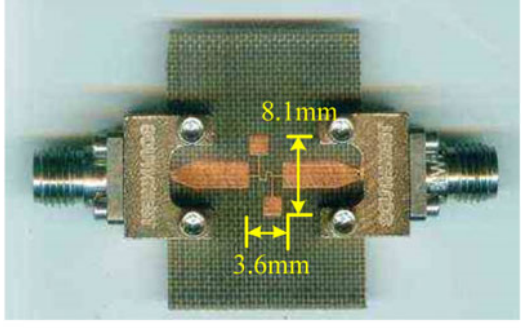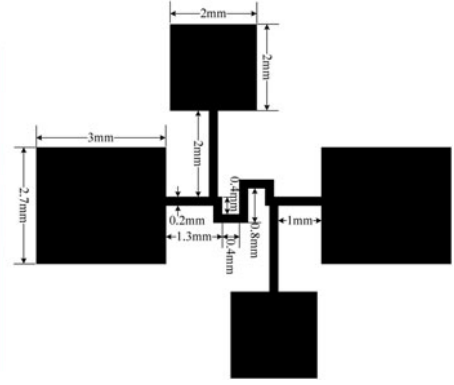

**Fig. 2.** The proposed harmonic rejection filter.

## B) Impedance matching

An HSMS 286 Schottky diode is applied in the proposed rectifier, which shows capacitive impedance at the operating frequency and input microwave power level. Therefore, an impedance transformer is required to match the complex impedance to 50 $\Omega$.

In the proposed design, the large signal model of the HSMS 286 Schottky diode is applied. The impedance of the diode [10] is defined as

$$Z_d = \frac{\pi R_s}{\cos\theta_{on}\left(\dfrac{\theta_{on}}{\cos\theta_{on}} - \sin\theta_{on}\right) + j\omega R_s C_j\left(\dfrac{\pi - \theta_{on}}{\cos\theta_{on}} + \sin\theta_{on}\right)}, \quad (1)$$

where $C_j$, $R_s$, and $\theta_{on}$ are the junction capacitance, the series resistance, and the forward-bias turn-on angle, respectively. A $\lambda_g/8$ short-ended microstrip line is equivalent to an inductance at the fundamental frequency. Therefore, this microstrip line cancels the imaginary part of diode impedance to enhance the rectifying efficiency at the fundamental frequency [9]. The inductive reactance of the microstrip line is defined as

$$jX_L = j2\pi f_o L = jZ_o \tan(\beta l), \quad (2)$$

where $Z_o$ and $\beta$ are the characteristic impedance and propagation constant, respectively. The equivalent inductance is $L = Z_o/2\pi f_o$, while $l = \lambda_g/8$. At the second harmonic, the input of the short-ended microstrip line is open which has no effect to the impedance matching circuit.

The MW-to-DC conversion efficiency is greatly dependent on the length of the parallel microstrip line. Figure 4 shows the relation between the conversion efficiency and the length of the parallel microstrip line with the optimum load resistance of the rectifier. In Fig. 4, the maximum conversion efficiency is 76%, when the line's length is 5.32 mm which approximates to $\lambda_g/8$. The MW-to-DC conversion efficiency drops obviously while the length deviates from $\lambda_g/8$. In the simulated results, the curve is not smooth due to an insufficient number of interpolation which is automatically controlled by ADS software.

When the diode self-bias DC voltage $V_o$ equals to half-diode reverse break-down voltage $V_{br}$, the conversion efficiency achieves the maximum value [11, 12]. The complex impedance of the diode HSMS-286 with different DC load resistances are simulated, and the results are shown in Fig. 5 when $V_o = 3.7$ V. The input impedance of the HSMS-286 is equal to 126 $\Omega$ at a 120 $\Omega$ DC load.

Figure 6 shows the simulated results of the impedance matching circuit used in the proposed rectifier. The matching circuit contains the parallel $\lambda_g/8$ short-ended microstrip line and two short series microstrip lines. At 5.8 GHz, the return frequency and an open circuit after $\lambda_g/4$ microstrip line. Normally, a chip capacitor cannot produce zero impedance since its capacitance is not infinity and the loss cannot be neglected at microwave frequency. Thus, an output low-pass filter is proposed to replace the chip capacitor and $\lambda_g/4$ microstrip line in schematic, which rejects the fundamental frequency microwave and its harmonics. Figure 7 shows the detailed microstrip structure of the low-pass filter which realized by the artificial transmission lines [13]. The interdigital structured microstrip is used to suppress the third harmonic.

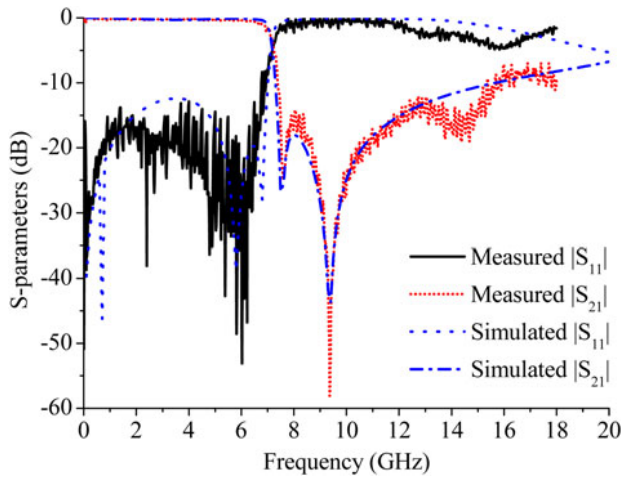

**Fig. 3.** The simulated and measured results of the low-pass filter.

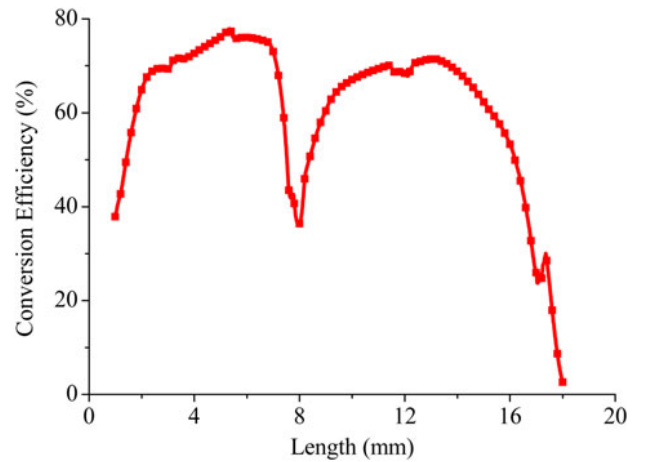

**Fig. 4.** MW-to-DC conversion efficiency according to the length of the parallel microstrip line.

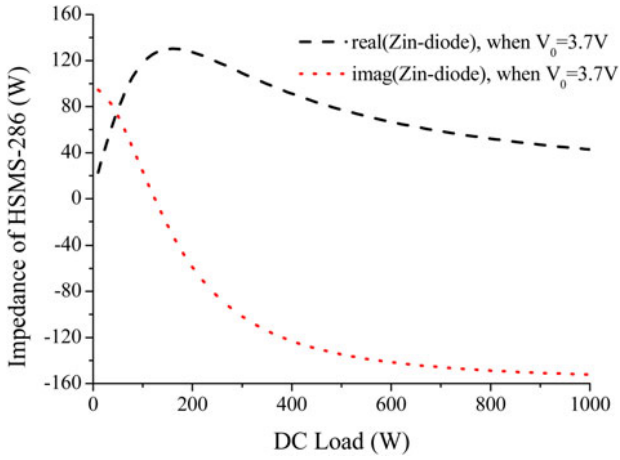

**Fig. 5.** Complex impedance of an HSMS-286 Schottky diode with various load resistances at 5.8 GHz.

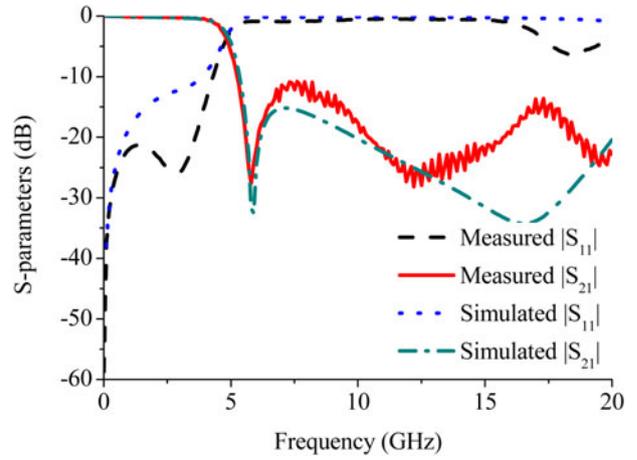

**Fig. 8.** Reflection and transmission characteristics of the low-pass filter.

The reflection and transmission characteristics of the applied low-pass filter are shown in Fig. 8. The reflection and the transmission coefficients at the fundamental frequency of 5.8 GHz are −0.8 and −27.4 dB, respectively. Furthermore, the reflection and the transmission characteristics are −0.46 and −25.6 dB at the second harmonic, −4.1 and −16.6 dB at the third harmonic, respectively. The low-pass filter obtains a good performance of preventing the fundamental frequency microwave and its higher order harmonics from leaking to the DC output port. Then, the harmonics are reflected back to the diode for power recycling, which may improve the DC-to-MW rectifying efficiency.

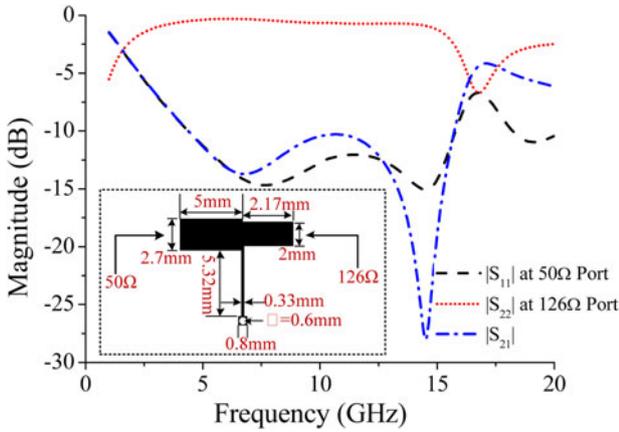

**Fig. 6.** Simulated insertion and return loss of the impedance matching circuit.

loss on both input filter and diode port is greater than 13 dB, and the simulated insertion loss is 0.3 dB.

## C) Output low-pass filter

Conventional output filter of the rectifier is composed of a $\lambda_g/4$ microstrip line and a lumped capacitor. The capacitor is assumed to produce a short circuit at the fundamental

## III. EXPERIMENTAL RESULTS

The rectifying circuit is realized on a F4B-2 substrate with $\varepsilon_r = 2.65$ and thickness of 1 mm. Figures 9 and 10 show the photo and the three-dimensional (3D) structure of the proposed 5.8 GHz microwave rectifier without any lumped passive components, respectively. The presented design applies two low-pass filters as discussed in the above to replace the lumped capacitors at the input and output in conventional microwave rectifiers.

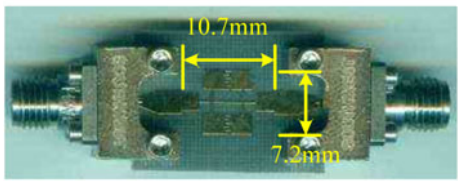

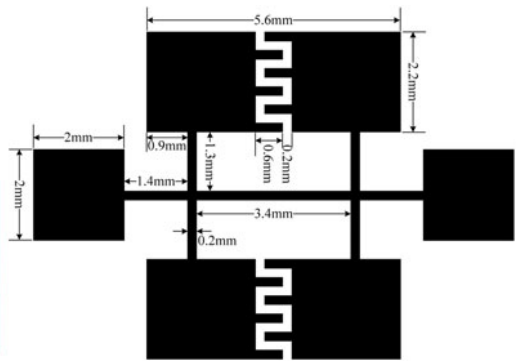

**Fig. 7.** Microstrip structure of the output low-pass filter.

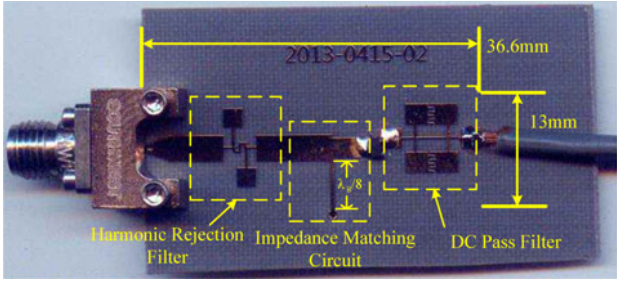

Fig. 9. The fabricated microwave rectifier.

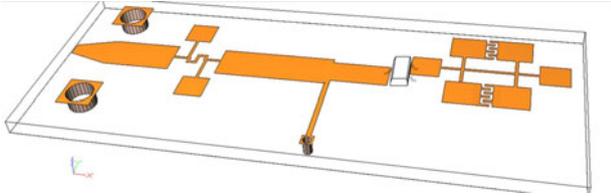

Fig. 10. 3D structure of the rectifier.

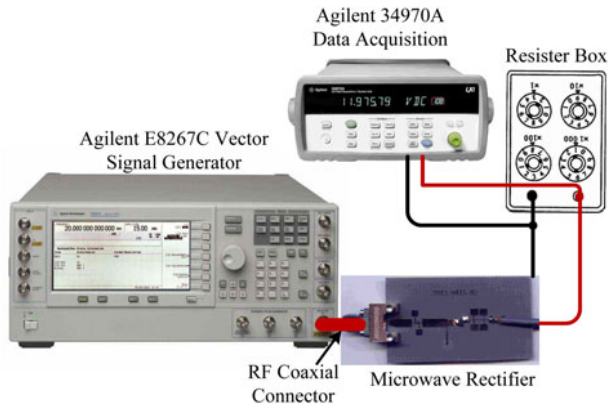

Fig. 11. The measurement system of the rectifier.

Fig. 11 shows the measurement system of the proposed rectifier. In this system, an Agilent E8267C vector signal generator is used as a microwave source which provides the power level from −130 to 20 dBm. The DC output voltage of the resister box is measured by an Agilent 34970A Data Acquisition.

The simulated and measured DC output voltage and MW-to-DC conversion efficiency at 18 dBm microwave input power are shown in Fig. 12. The lines marked with triangle and rectangle show the output voltage and the conversion efficiency, respectively. The measurements agree well to the simulated results. At 18 dBm microwave input power, the maximum MW-to-DC conversion efficiency of the proposed filter achieves 67.5%, when the DC output voltage reaches 2.92 V with a 200 Ω DC load. If the input filter and the matching circuit are de-embedded, the MW-to-DC conversion efficiency is equal to 74.2%.

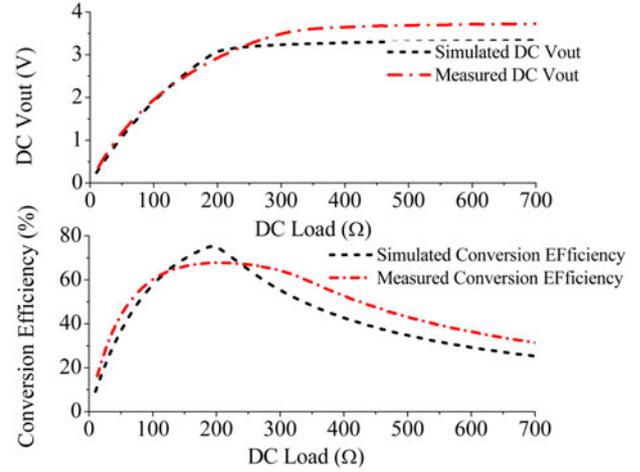

Fig. 12. Simulated and measured DC output voltage and MW-to-DC conversion efficiency at 18 dBm.

Figure 13 shows the measured MW-to-DC efficiency of the rectifier at various microwave input power level. Its highest MW-to-DC efficiency reaches 68.1% at a microwave input power of 19 dBm and a DC load of 190 Ω. When the input MW power increases, the optimal DC load decreases to maintain a constant DC voltage to achieve the best MW-to-DC conversion efficiency. The measured output voltage is higher than the simulation when the DC load is high, since the diode model is possibly not so accurate under high DC bias voltage. The highest MW-to-DC efficiency is higher than 60% when the input microwave power varies from 13 to 20 dBm. Figure 14 shows the measured MW-to-DC efficiency dependent on frequency. The variation of frequency results in an impedance mismatching, and leads to the drop of the conversion efficiency.

## IV. CONCLUSION

In this paper, we investigated a 5.8 GHz microwave rectifier without any passive lumped components to achieve a high

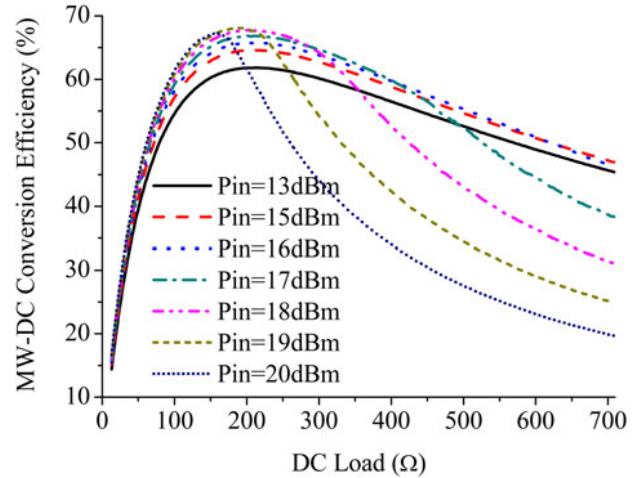

Fig. 13. Measured MW-to-DC conversion efficiency with respect to the load at different input power.

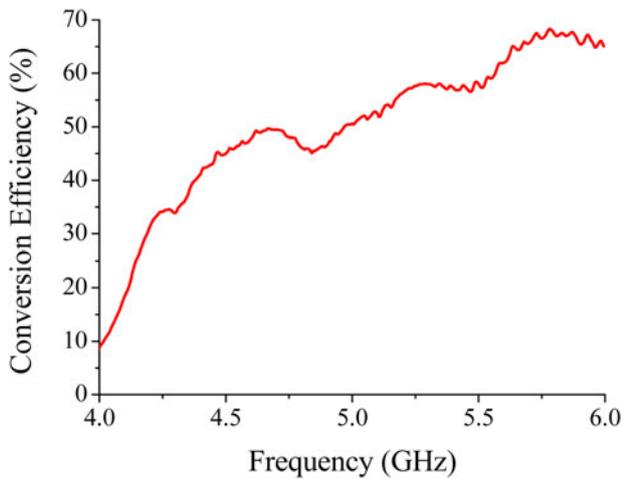

**Fig. 14.** The conversion efficiency with respect to the frequency at a 19 dBm input power and a 200 Ω load.

MW-to-DC conversion efficiency. In the design, the conversion efficiency is widely varied dependent on the lengths of the parallel microstrip transmission line, which implies that impedance matching is critical to achieve a high MW-to-DC efficiency. Two microstrip low-pass filters are applied to replacing the lumped capacitors in the input and output of a conventional rectifier to enhance the conversion efficiency. The proposed rectifier shows that a microwave rectifier without any capacitors may obtain a high efficiency and be applied for higher microwave bands. Compared with the rectifiers with lumped elements [14–16], this design is conducive to enhancing the consistency of the circuit in volume manufacturing.

This rectifying circuit realized a MW-to-DC conversion efficiency of 68.1% at 19 dBm input microwave power, and its best conversion efficiency keeps greater than 60% with the input microwave power from 13 to 20 dBm. This design may be conveniently applied to high-frequency wireless MPT systems.

## ACKNOWLEDGEMENTS

This work was supported in part by the NSFC 0971051, 973 program 2013CB328902, and NCET-12-0383.

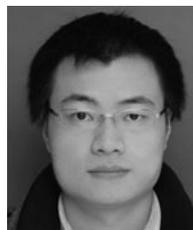

**Biao Zhang** was born in Xi'an, China in September 1987. He received his B.Sc. degree in Electronics Information Engineering from Sichuan University of China in 2009. He is currently working toward his Ph.D. degree in Communication and Information System in Sichuan University. His main research interests are the design of microwave passive circuits and microwave power transmission systems.

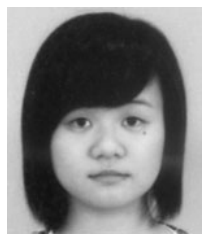

**Wan Jiang** was born in Chengdu, China in February 1989. She received her B.Sc. degree in Electronics Information Engineering from Sichuan University of China in 2011. Now, She is a Ph.D. candidate of Sichuan University. She is involved in the research of frequency-scan antennas and microwave passive circuits.

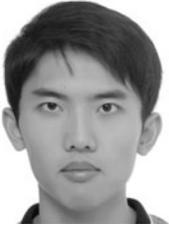 **Chengyang Yu** was born in Bozhou, China in August 1988. He received his B.Sc. degree in Electronics Information Engineering from Sichuan University of China in 2010. Now, he is a Ph.D. candidate of Sichuan University. He is involved in the research of rectennas and microwave passive circuits.

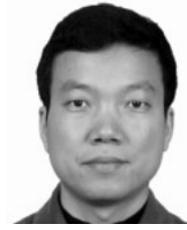 **Changjun Liu** was born in Xingtai, China in April 1973. He received his B.Sc. degree in Applied Physics from Hebei University of China in 1994, and M.Sc. majoring in Radio Physics in 1997 and Ph.D. in Biomedical Engineering in 2000 from Sichuan University. Now, He is focusing on microwave/radio frequency circuits and microwave wireless power transmission systems.